\documentclass{ws-ijmpb}

\newcommand{\mb}[1]{ { \mbox{\boldmath{$#1$}}}  }
\newcommand{\mbs}[1]{ {\scriptsize \mbox{\boldmath{$#1$}}}  }
\newcommand{\sub}[2]{ \mbox{$#1$}_{\mbox{\scriptsize\boldmath{$#2$}}}}
\begin{document}

\markboth{Grzegorz Litak}
{
Spatial Fluctuations of the Order  Parameter in
$P$--Wave Superconductors}

\catchline{}{}{}

\title{
Spatial Fluctuations of the Order  Parameter in
a Disordered Superconductor with $P$-- Wave Pairing}
  
\author{Grzegorz Litak}

\address{
Department of Mechanics, Technical University of
Lublin \\ Nadbystrzycka 36, PL--20-618  Lublin, Poland.}

\maketitle
\pub{Received (\today)}{}
%{Revised (revised date)}

\begin{abstract}
\noindent The effect of nonmagnetic disorder 
on the
pairing amplitude is studied 
by means of a perturbation
method.
Using an extended one band Hubbard model with an intersite attraction
we analyze various solutions of $p$--wave pairing symmetry
and  discuss their
instability calculating 
 fluctuations of order parameter. The model is applied to describe the
effect of disorder in Sr$_2$RuO$_4$. The results shows that the real
solution with line nodes can be favoured by disorder.
\end{abstract}

\section{Introduction}

The recent discoveries of  superconductivity in 
Sr$_2$RuO$_4$\cite{Mae01,Agt97} 
and  UPt$_3$\cite{Gra00a,deV00}, apparently $p$--wave,  raised the
questions of their 
unconventional properties.
Particularly the  question of system response on nonmagnetic disorder 
attracted interest of many 
researchers\cite{deV00,Miy99,Mac98,Min01,Zur01,Lit01a,Mak00,Gra99} . It is clear 
that this effect can distinguish the unconventional superconductors\cite{Mar99,Bor94}.     
For  conventional superconductors (with isotropic s-wave paring and a 
large
coherence length) 
it is known, due to Anderson\cite{And59,Mak69,Gyo97}
that 
the non-magnetic disorder have negligible effect on 
superconducting gap and the critical temperature $T_C$. 

The fluctuations of the pairing amplitude are scaled as the inverse of coherence length
square ($\sim \xi^{-2}$)\cite{Gyo97}. 
In an analogy to that, for anisotropic superconductors,
disorder induces the effect of spatial 
fluctuations in the amplitude $|\Delta(ij)|$\cite{Lit01b,Lit01d} which can
destroy the superconducting phase.

In this paper we will consider the effect of disorder on various possible
$p$--wave solutions.
It is  connected with a dilemma\cite{Mae01}
appeared
recently  in context of  superconductivity in Sr$_2$RuO$_4$. Namely, the
question if superconducting order parameter posses or not the line nodes
in
the gap was asked. As among possible $p$--wave paired solutions there are
those with line nodes and without them, therefore it is important to
check if disorder can favour any of possible $p$--wave solutions.

For simplicity we are going to use the one band model\cite{Mic88,Lit01d,Ere01}.
 Such choice
can be
justified  knowing that one of the  promising model of
$p$--wave superconductibity  in Sr$_2$RuO$_4$ is based on the important
assumption that $\gamma$ band can be dominant as far as the appearance of
superconductivity is concerned while the electrons in other two
passive bands
($\alpha$ and $\beta$)
are paired because of proximity effect\cite{Zhi01}.

The paper is organized as follows. After short introduction in the
present section we present  the model in
the
clean limit and  introduce the Hamiltonian and approximations used in
paper (Sec. 2). Here we discuss the results of paring parameter 
obtained
for triplet superconductor.
Section  3  is devoted to a disordered p-wave superconductor. We include
diagonal nonmagnetic disorder and analyze
spatial fluctuations of
a pairing
potential. 
Particularly we calculate
standard deviations random site energies  $<\delta
\varepsilon_{i}>$ and for pairing potentials for neighbour sites $(ij)$
$<\delta
\Delta_{ij}>$. Their ratio $\Gamma_{ij}= <\delta
\Delta_{ij}>/<\delta
\varepsilon_{i}>$ can be regarded a criterion of pairing potential
fluctuations\cite{Lit01b,Lit01d}
Finally we investigate the  stability of
various solutions in
presence of disorder.
Section 5 contains conclusions and remarks.

\section{ The model.
}

We start a single band, extended, Hubbard model with intersite
attraction. It is defined by the following 
Hamiltonian\cite{Mic88}:

%1
\begin{equation}
H=  \sum_{ij \sigma} (\varepsilon_i \delta_{ij} +t_{ij}
)c_{i\sigma}^{+}c_{j\sigma} + \sum_{i j \sigma \sigma'} 
\frac{W_{ij}}{2} n_{i \sigma} n_{j \sigma'}
- \mu \sum_{i \sigma}
c_{i\sigma}^+c_{i\sigma}~,
\label{eq1}
 \end{equation}
where $c_{i\sigma}^+$ and $c_{i \sigma}$ are usual, fermionic
operators    which     create     and 
annihilate, respectively, an electron 
with spin
$\sigma$  at the
lattice
site labeled by $i$, $t_{ij}$ is a electron hopping integral, $n_{i \sigma}$
is the operator of particle number  of spin $\sigma$ at site $i$,
$\varepsilon_i$ is the 
site
energy, varying from site to site 
in random fashion,  at the site $i$ with mean value
$\varepsilon_0=<\varepsilon_i>=0$ and $W_{ij}$ is the 
interaction potential
of two electrons with opposite spins on neighbour sites $i,j$. $\mu$ 
denotes the chemical
 potential.

Hartree--Fock--Gorkov equation of motion for matrix Greens
$4 \times
4$ functions ${\mb G}(i,j;\omega)$ yields:

%2
\begin{equation}
\label{eq2}
\sum_l \left( \begin{array}{c} \left[(\imath \omega_n -\varepsilon_i + \mu)
\delta_{il}+t_{il}\right] \mb 1~~~~~ \mb \Delta_{il} \\
 \mb \Delta_{il}^+~~~~~ \left[(\imath \omega_n +\varepsilon_i - \mu) \delta_{il}
-t_{il} \right] \mb 1  \end{array} \right)
 \left( \begin{array}{c} \mb G_{11}(l,j;\imath \omega_n)~~
\mb G_{12}(l,j;\imath \omega_n) \\
\mb G_{21}(l,j;\imath \omega_n)~~ \mb G_{22}(l,j;\imath \omega_n) \end{array}
\right)
=
\delta_{ij} \mb 1
\end{equation}
with the spin dependent 4$\times$4  Green function. Each of  its components
$\mb G_{nm}(l,j;\imath
\omega_n)$ is defined as:
%3
\begin{equation}
\label{eq3}
\mb G_{nm}(l,j;\imath \omega_{\nu})= \left(\begin{array}{cc} G_{nm}^{\uparrow
\uparrow}(l,j;\imath \omega_{\nu}) &
G_{nm}^{\uparrow
\downarrow}(l,j;\imath \omega_{\nu}) \\ G_{nm}^{\downarrow
\uparrow}(l,j;\imath \omega_{\nu}) &
G_{nm}^{\downarrow
\downarrow}(l,j;\imath
\omega_{\nu})
\end{array} \right),~~~~~n,m=1,2~.
\end{equation}

The order parameter for triplet paring reads:
%4
\begin{equation}
\label{eq4}
\mb \Delta_{ij} =
\left(
\begin{array}{cc}
\Delta_{ij}^{\uparrow \uparrow} & \Delta_{ij}^{\uparrow \downarrow} \\
\Delta_{ij}^{\downarrow \uparrow} & \Delta_{ij}^{\downarrow \downarrow}
\end{array}
\right).
\end{equation}

The gap equation for p-wave order parameter:

%5
\begin{equation}
\label{eq5} 
\Delta_{ij}^{\alpha \alpha'} = U_{ij} \frac{1}{\beta} \sum_n {\rm
e}^{\imath 
\omega_n
\eta}
G_{12}^{\alpha \alpha'} (i,j;\imath \omega_n)
,~~~~~ \alpha, \alpha'
=\uparrow, \downarrow~,
\end{equation}
where $G_{12}^{\alpha \alpha'}$ is a matrix element of 
$4 \times
4$ functions Green functions.

The above
equation Eq. \ref{eq5}  have to be
completed by 
the
 corresponding equations for the chemical potential
$\mu$ that
satisfies the   
following relation:
%6
\begin{equation}
\label{eq6}
n= \frac{2}{\beta} \sum_n {\rm e}^{\imath
\omega_n \eta}
G_{11}^{\uparrow \uparrow}(i,i;\imath \omega_n),
\end{equation}  
where $n$ is the number of electrons per unit cell.

%fig1
\begin{figure}[htb]
\leavevmode
\hspace*{0.5cm}
\epsfxsize=4.5cm
\epsffile{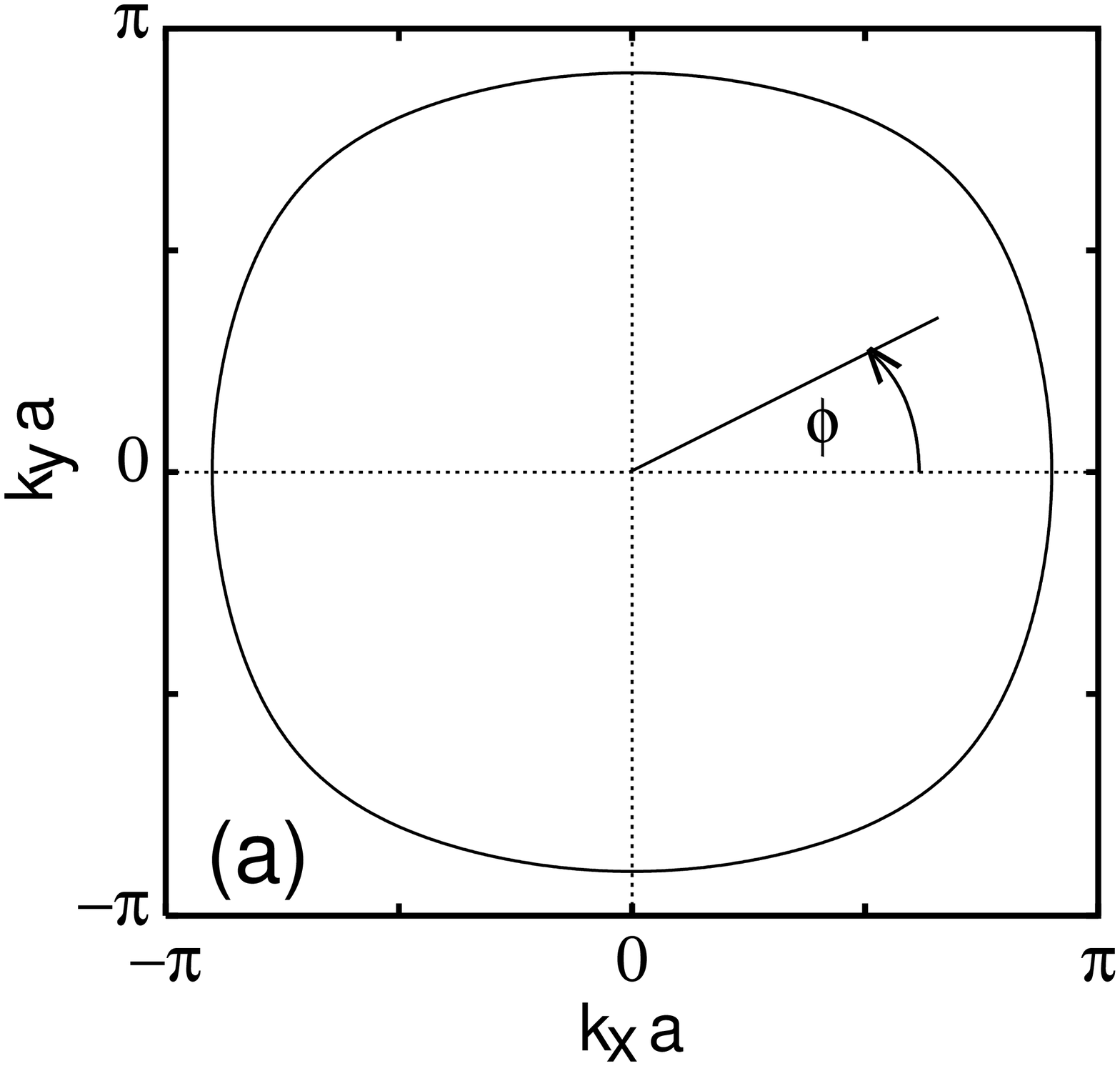}
\hspace*{1cm}
\vspace*{1.3cm}
\epsfxsize=4.5cm
\epsffile{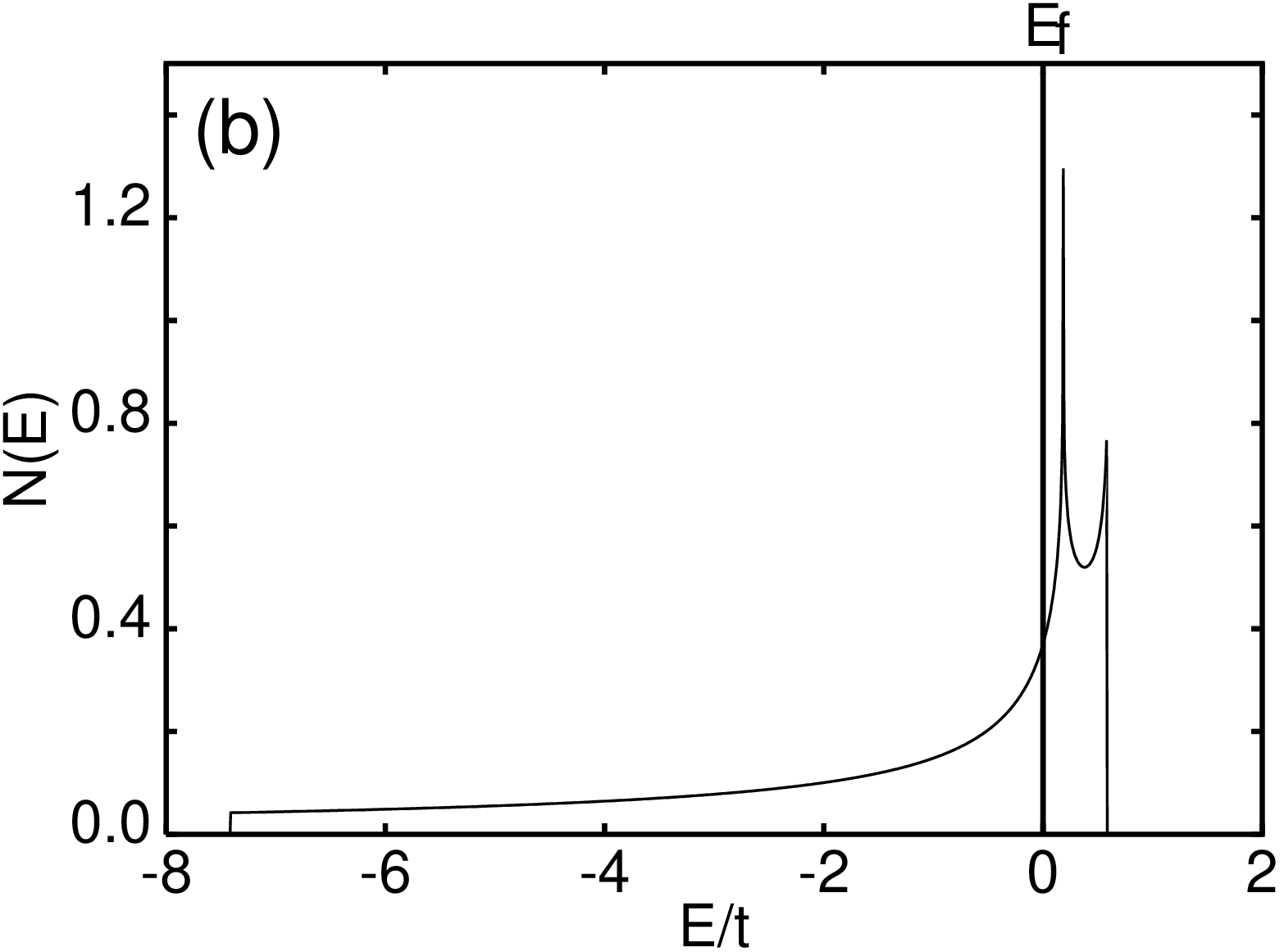}
\vspace{-1cm}
\caption{The Fermi surface (a) and the density of states $N(E)$ (b) for
$\gamma$
band Sr$_2$RuO$_4$ ($t'=0.45t$, $n=1.34$).}
\end{figure}

\section{$P$--wave solutions for a clean system at $T=O$K}

Taking the Fourier transform of the Green function for a clean system
($\varepsilon_i=0$) we can write its equation of motion:

%7
\begin{equation}
\left[ \begin{array}{cc} (\omega-\sub{\epsilon}{k}  +\mu) \mb 1    &
\sub{\mb \Delta}{k}
 \\
\sub{\mb \Delta}{k}^* &   (\omega
+\sub{\epsilon}{ k} -\mu) \mb 1   \end{array}      \right] {\mb G}^0(\mb
k; \omega)={\mb 1}
\label{eq7}.
\end{equation}

Here, we  assumed that the electron hopping  integrals $t_{ij}$ can take a
nonzero values 
for
the  nearest
and
next nearest neighbour sites  only. For a clean system $t_{ij}$ can be
expressed
in
$\mb k$-space: 
$\sub{\epsilon}{k}= \sum_j t_{ij} {\rm e}^{-\imath \mbs R_{ij} \mbs k}$
as:
%8
\begin{equation}
\label{eq8}
\sub{\epsilon}{k} = -2 t (\cos k_x +
\cos k_y)-4t'\cos k_x \cos k_y~,
\end{equation}
where $t$ represents the nearest neighbour site amplitude of electron
hopping, while $t'$ corresponds to next nearest neighbour
one.

\noindent As can be readily shown 
the $p$--wave superconducting order parameter ${\mb \Delta}_{\mbs k}$ 
can be written as
%9
\begin{equation}
\sub{\mb \Delta}{k}= \mb \Delta_{x} \sin k_x +\mb \Delta_{y} \sin k_y,
\label{eq9} 
\end{equation}  
where $\mb \Delta_{x}$ and $\mb \Delta_{y}$ are complex $2 \times 2$
matrix of pairing potentials. 
Assuming the order parameter 
$\sub{\mb \Delta}{k}=
 \imath \hat{\sigma}_y\hat{{\mathbf \sigma}}\cdot{\bf d}({\bf k})$,
with ${\bf d}({\bf k})=(0,0,d^z({\bf k}))$\cite{Lit01c}
and
 $d^z({\bf k})=\sub{\Delta}{k}$   
 The  gap
 equation can be written as\cite{Mic88}:
%10
\begin{equation}
\label{eq10}
\sub{\Delta}{k} = 
  \frac{1}{N} \sum_{\mbs q} \frac{\sub{W}{kq} \sub{\Delta}{q}}
{ 2 \sub{E}{q}} {\rm Tanh} \left( \frac{\beta \omega}{2} \right)~,
\end{equation}
where for $p$--wave pairing we can assume
%11
\begin{equation}
\label{eq11} 
\sub{W}{kq} 
 = |W| \left( 
2 {\rm sin}(k_x) {\rm sin}(q_x) + 2 {\rm sin}(k_y) {\rm sin}(q_y)
\right)
\end{equation}
and $\sub{E}{q}$ denotes quasi-particle energy: 

%12
\begin{equation}
\label{eq12} 
\sub{E}{q} = \sqrt{\sub{\tilde{\epsilon}}{q}^2 -\sub{\Delta}{q}^2}~~,~~~~~~ 
\sub{\tilde{\epsilon}}{q}= \sub{\epsilon}{q} -\mu.
\end{equation}
and the equation for chemical potential:
%13
\begin{equation}
\label{eq13} 
n-1= -\frac{2}{N} \sum_q \frac{\sub{\tilde{\epsilon}}{q}}{ 2
\sub{E}{q}} 
~~{\rm Tanh} \left( \frac{\beta \sub{E}{q}}{2} \right) 
\end{equation}

The corresponding free energy $F$ for a finite temperature $T$ can be
calculated from the
following formula:

%14
\begin{equation}
\label{eq14}
F=\sum_{\mbs k} \left[ -(n-1) \sub{\tilde \epsilon}{k} -2k_BT~ {\rm ln}
\left( 2
\cosh
\frac{ \sub{E}{k} }{2k_BT} \right) -\frac{|\sub{\Delta}{k}(T)|^2
}{W} \right]
\end{equation}

To perform numerical calculations we have fitted our one band
system
parameters
to the realistic $\gamma$ band structure of 
Sr$_2$RuO$_4$\cite{Agt97,Lit01a,Lit01d,Lit01c,Mac96}.
Fig. 1 presents the corresponding Fermi surface (Fig. 1a) as well as the
electron density
of states $N(E)$ (Fig. 1b) of the normal state. 
Solving Eqs. \ref{eq10}-\ref{eq13} we have calculated  the pairing
parameter $\Delta_x$ and  $\Delta_y$ for various possible $p$--wave
superconducting states.
Namely, depending on relative values $\Delta_x$ and
$\Delta_y$: dipole ($\Delta_x \neq 0$ and  $\Delta_y = 0$),
real ($\Delta_x=\Delta_y$) and complex one ($\Delta_x=\imath
\Delta_y$)
where found.  
In Fig. 2a we plotted  $\Delta_x$ for these solutions
versus band filling $n$.
The lines are denoted by letters: D, R, C; respectively.

%fig2
\begin{figure}[htb]
\leavevmode

\vspace{-1cm}
\hspace*{0.5cm}
\epsfxsize=4.0cm
\epsffile{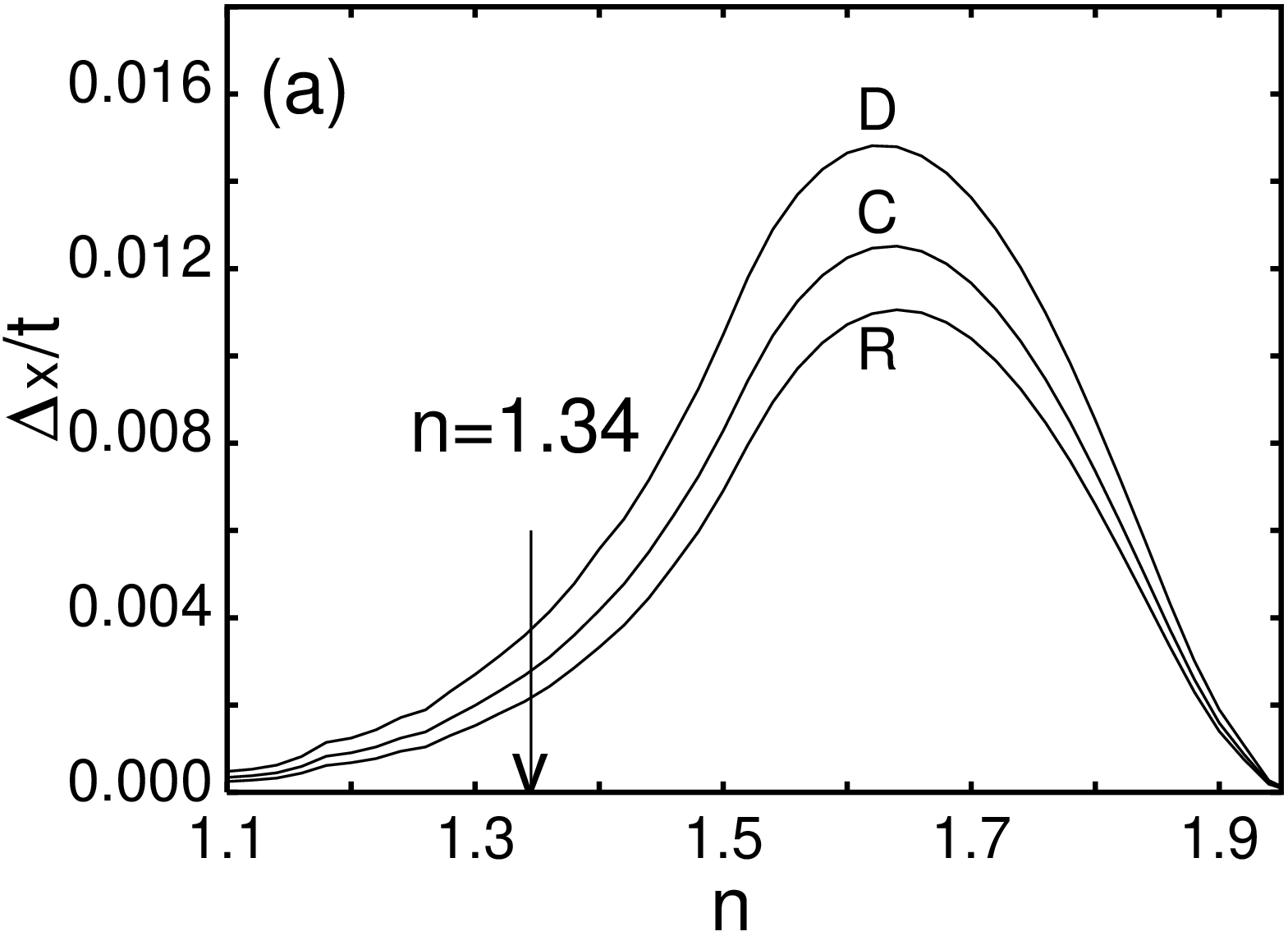}
\hspace*{2.0cm}
\epsfxsize=4.cm
\epsffile{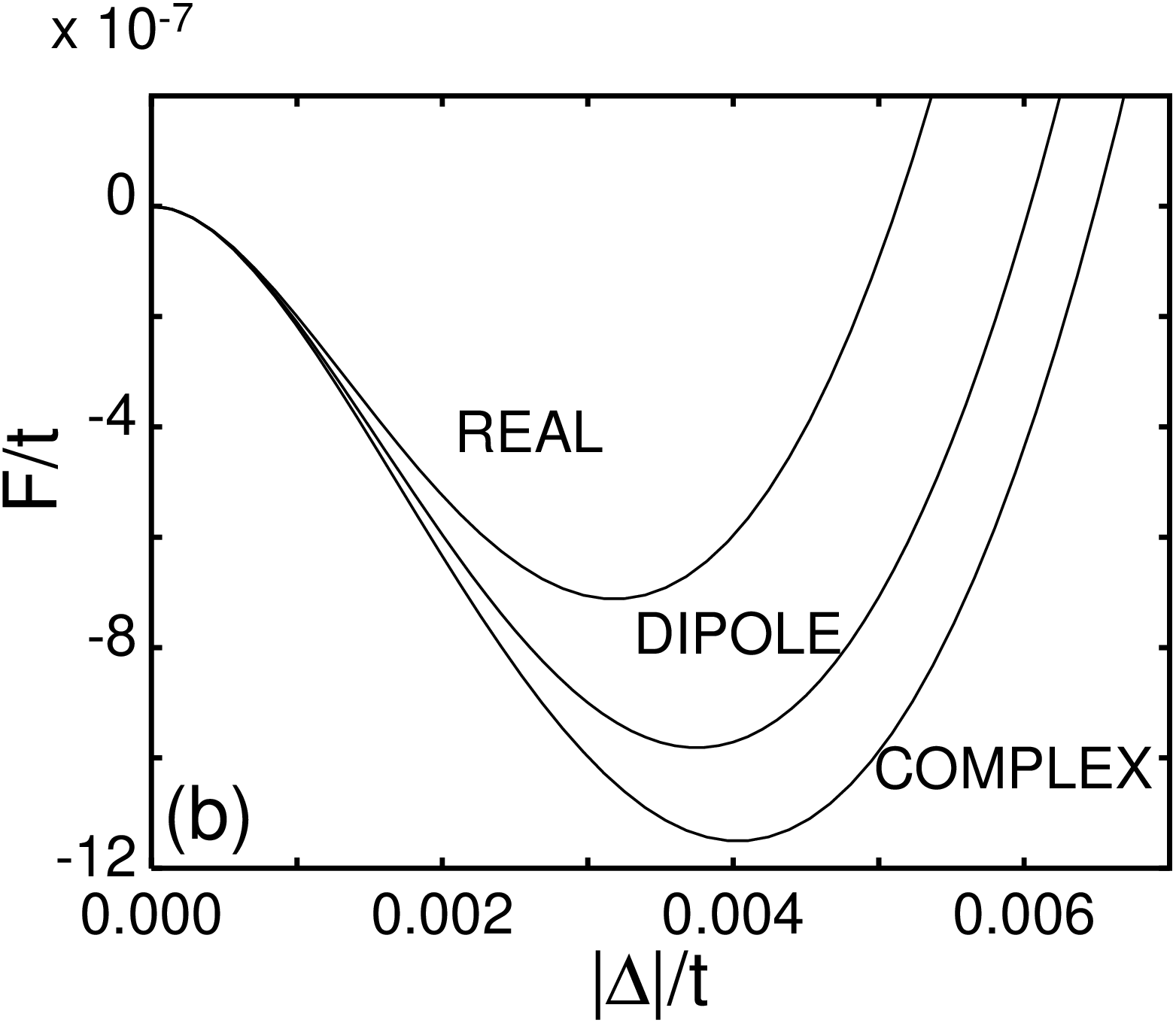}
\vspace{0cm} 
\caption{(a) the pairing parameter $\Delta_x$ versus band filling $n$ and
(b) the free energy $F$ as a function of $|\Delta|$ ($|\Delta|=
\sqrt{\Delta_x^2+\Delta_y^2}$) for different solutions with dipole (D)
real (R) 
and complex (C) order parameters, respectively. The intersite
attraction $W=-0.4605t$ and temperature  $T=0$K. The arrow in Fig. 2a
corresponds to the $\gamma$
band filling in Sr$_2$RuO$_4$ $n=1.34$.}
\end{figure}

%fig3
\begin{figure}[htb]
\leavevmode

\vspace{0cm}
\hspace*{0cm}
\epsfxsize=4.0cm
\epsffile{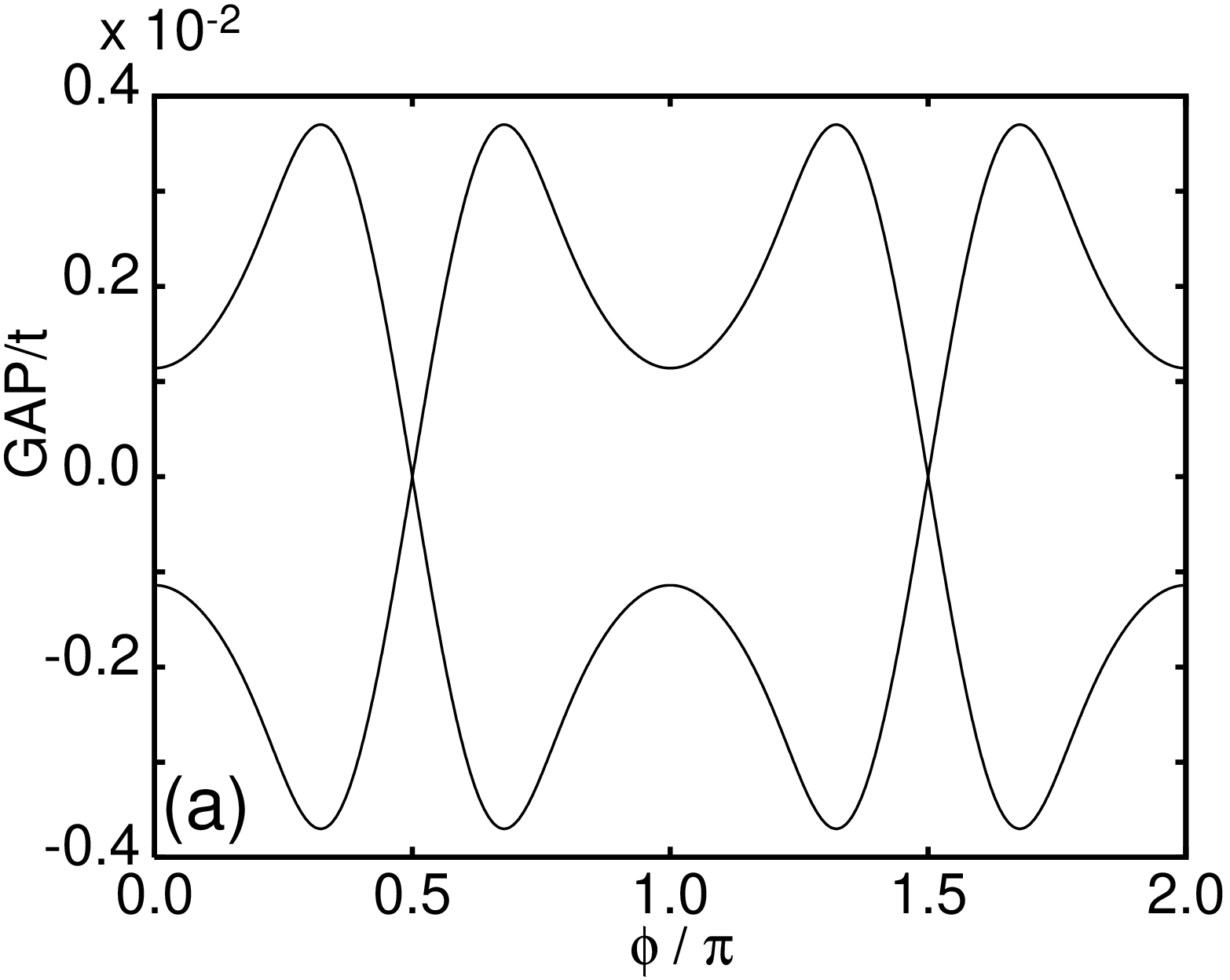}
\hspace*{2cm}
\epsfxsize=4.0cm
\epsffile{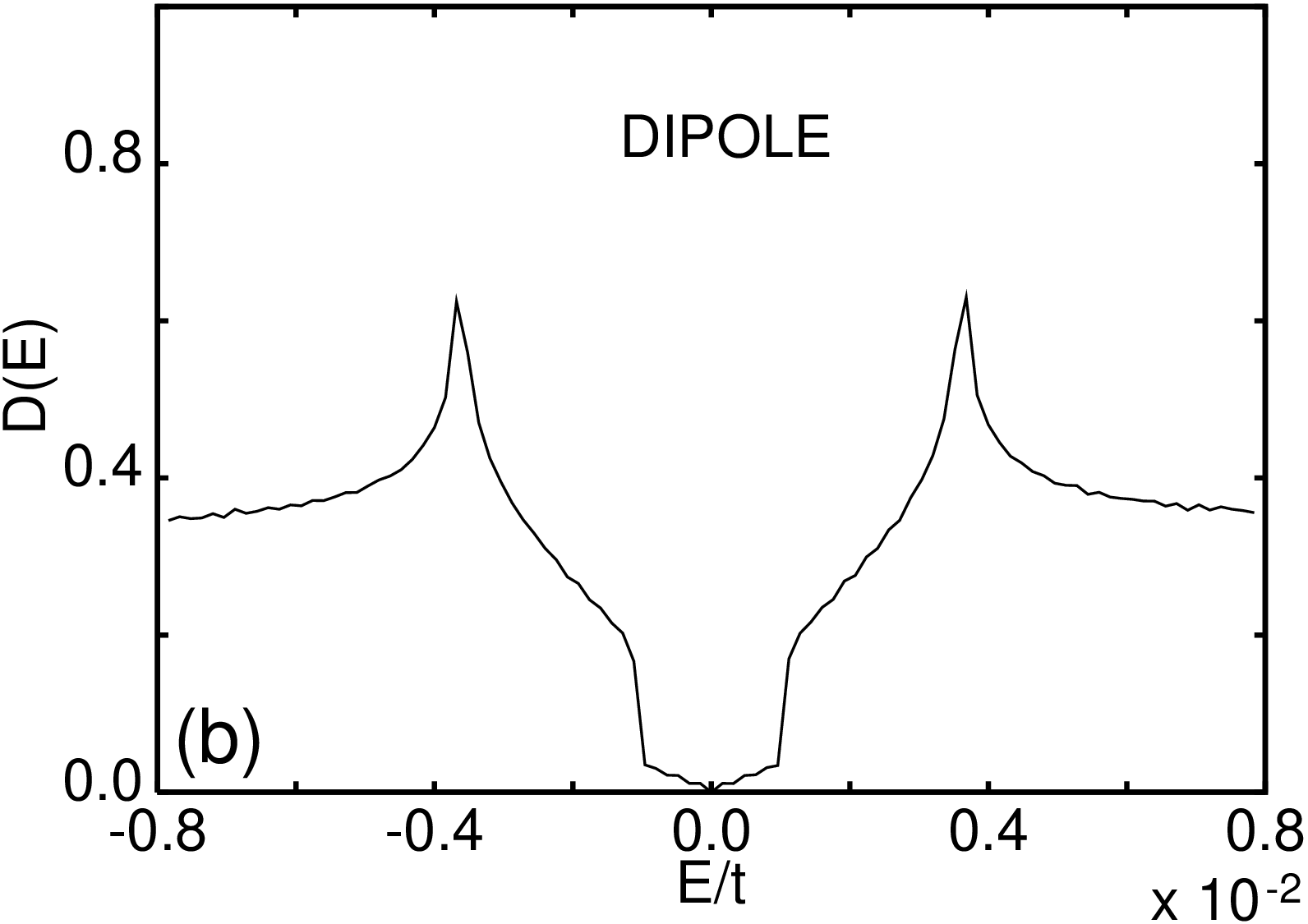}
\vspace*{-0.5cm}

\hspace*{0cm}
\epsfxsize=4.0cm
\epsffile{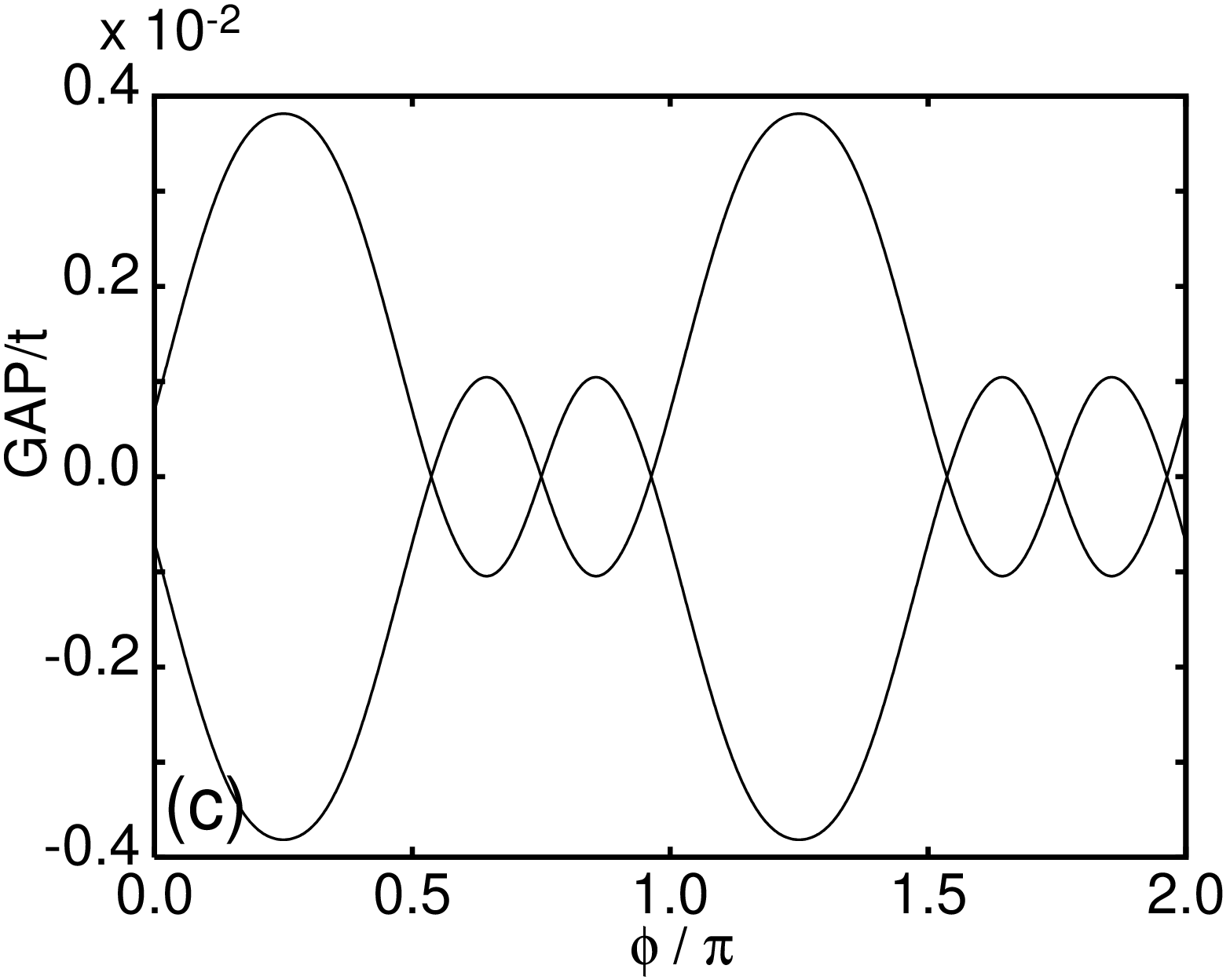}
\hspace*{2cm}
\epsfxsize=4.0cm 
\epsffile{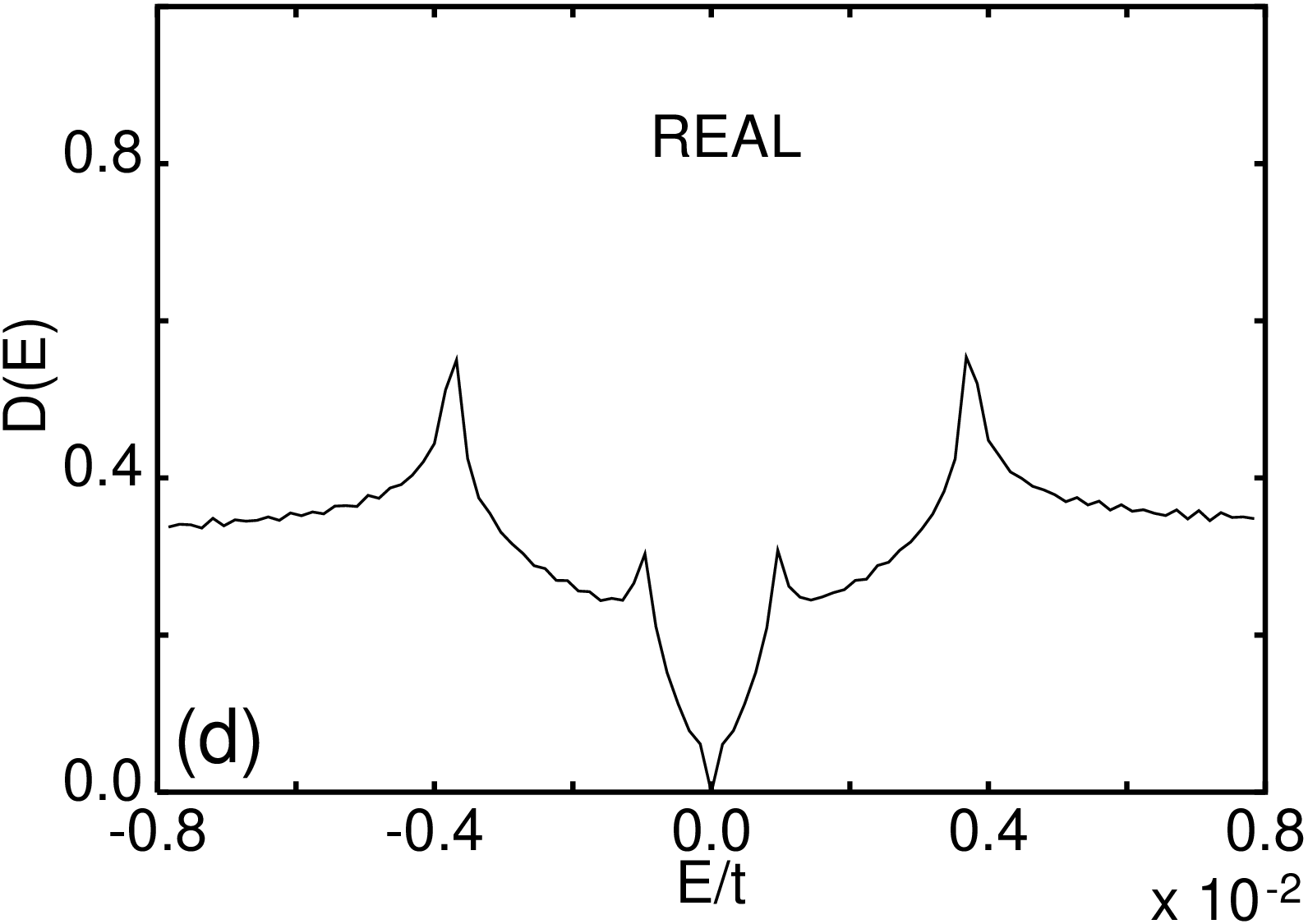}

\vspace*{-0.5cm}

\hspace*{0cm}
\epsfxsize=4.0cm
\epsffile{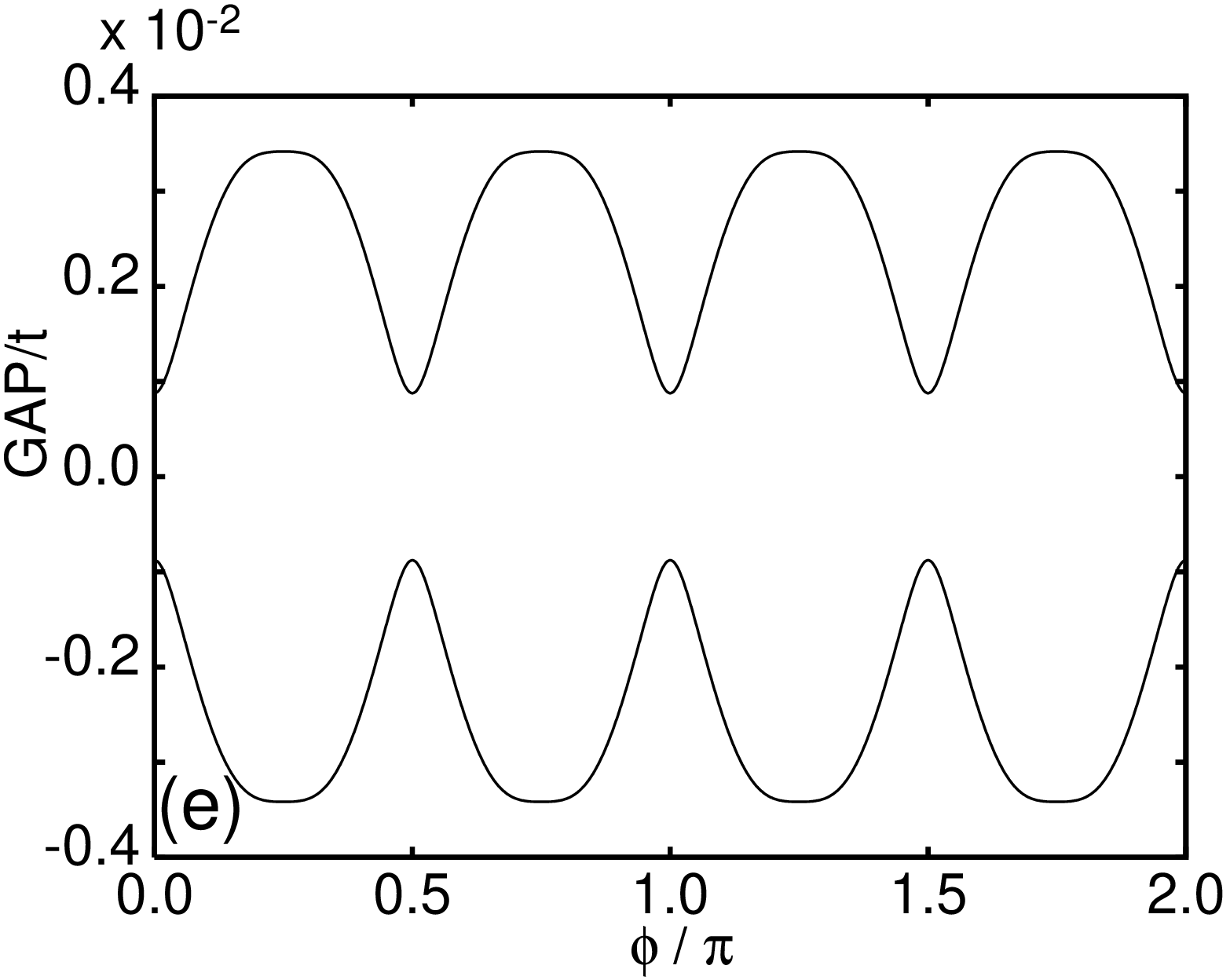}
\hspace*{2cm}
\epsfxsize=4.0cm
\epsffile{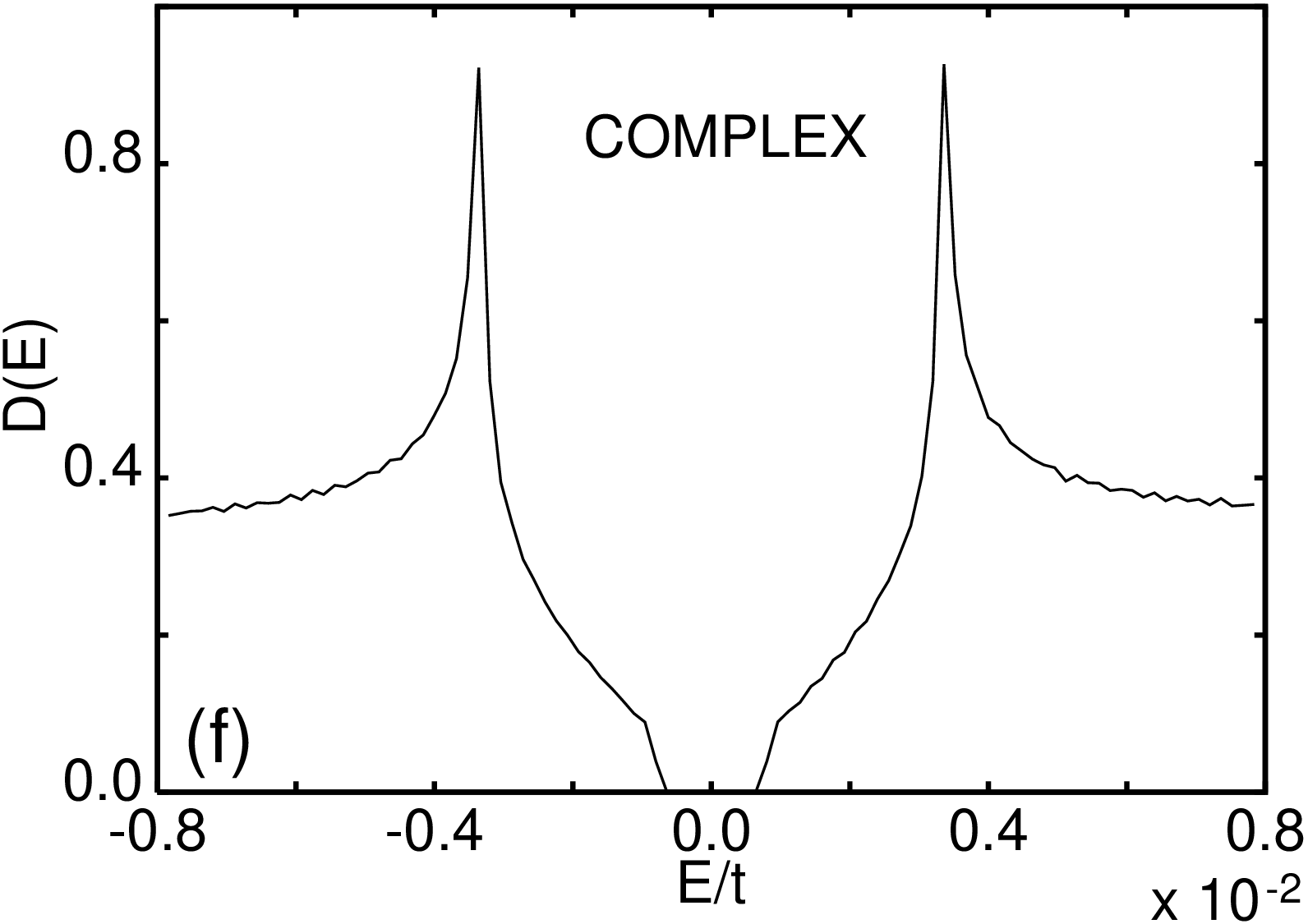}

%\vspace{2cm}
\caption{The angular gap ($\pm \sub{E}{k}(\sub{\epsilon}{k}=\mu)$)
dependence and the quasi particle density of states
for
different solutions: dipole (a,b) real (c,d)
and complex (e,f) type order parameter; respectively. The angle $\phi$ is
defined in
Fig. 1a.}
\end{figure}

%fig4
\begin{figure}[htb]
\leavevmode
\hspace*{2.5cm}
\epsfxsize=5.0cm
\epsffile{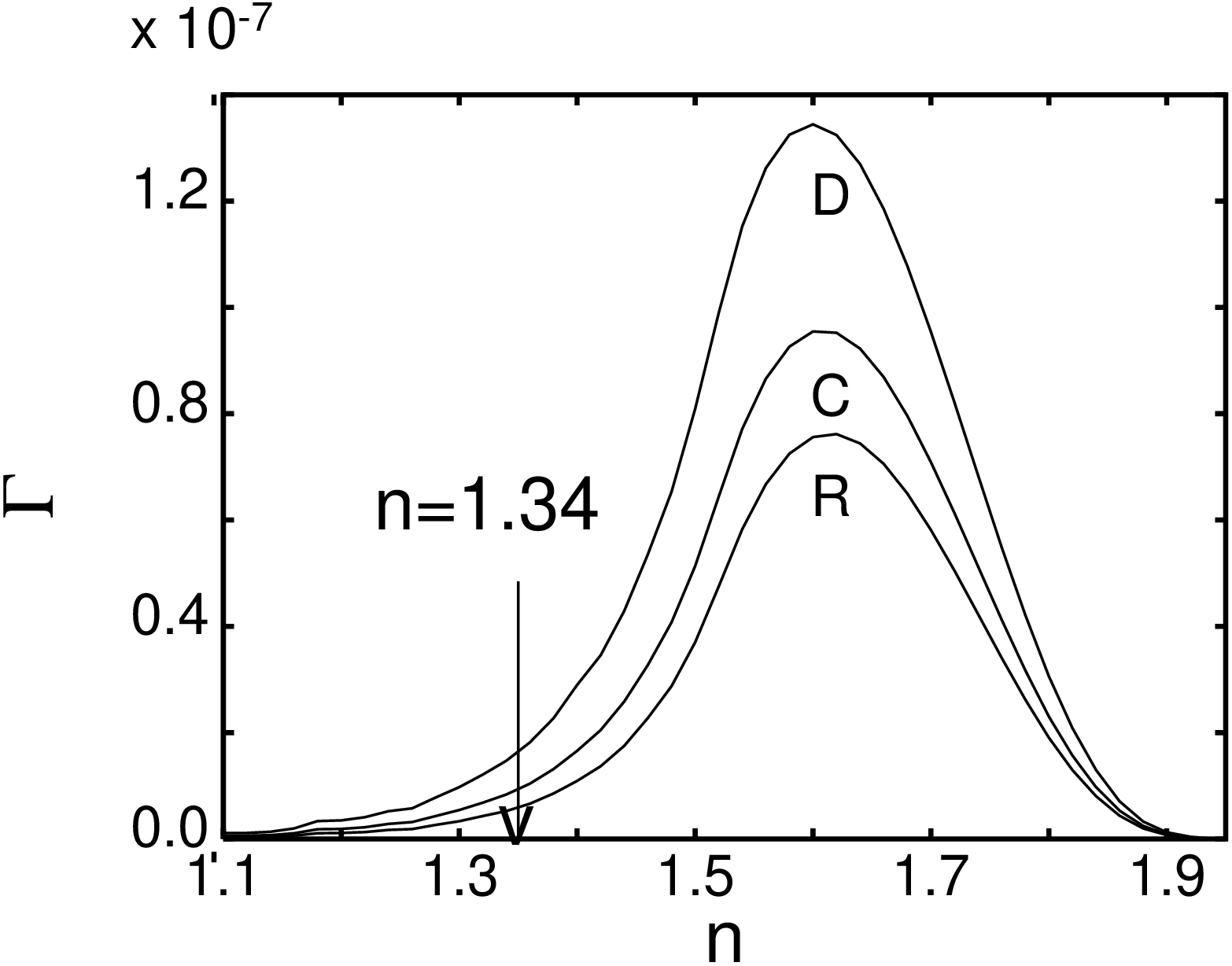}
\hspace*{2.0cm}
\vspace{-0.5cm}
\caption{Fluctuation parameter $\Gamma$ as a function of band filling
for different solutions: dipole (D) real
(R) and complex (C) order parameter, respectively. The intersite
attraction $W=-0.4605t$ and the band filling  $n=1.34$ correspond to
$\gamma$
band occupation
of Sr$_2$RuO$_4$.}
\end{figure}

The interaction parameter used in calculations ($W=-0.4605t$) were chosen
to
give the same $T_c=1.5$K for all $p$--wave solutions, as for the pure
Sr$_2$RuO$_4$.
Examining the free energy $F$ (Eq. \ref{eq14}) we have also found 
the energetic preferences of these three solutions.
In Fig. 2b the free energy $F$ was plotted as a function of $|\Delta|$.
In 
One can easily see that the complex solution reaches the global minimum of
free
energy $F$.
Additionally,
the angular dependences of gaps  ($\phi \in
[0,2\pi]$) have been plotted
in Figs. 3a, b and c.
Note that both dipole and real solutions posses line nodes while
the complex
one has a finite gap in any angle $\phi$ direction ($\phi$ is defined in
Fig 1a).  For clarity we have plotted corresponding quasi particle density
of states in Figs. 3b, d and f, respectively. Note that only in case with
the complex solution it has a  finite gap (Fig. 3f). Interestingly the
real solution has the largest amplitude (Figs. 3c, d).

\section{ Fluctuations of pairing potential.}  

In this section we investigate the stability of superconducting $p$--wave
states in presence of weak disorder.
Here, we apply the same strategy as in\cite{Gyo97,Lit01b,Lit01d} and we
treat random site energies $\varepsilon_i$ as
perturbations. 
To proceed we write  the Dyson equation for a Green
function $\mb G(i,j;\omega)$ evaluated at a frequency
$\omega$:

%15
\begin{equation}
\label{eq15}
{\mb G}(i,j;\omega) = {\mb G}^0(i,j;\omega) + \sum_l
{\mb G}^0(i,l;\omega) {\mb V}_l {\mb G}(l,j;\omega)~,
\end{equation}
where $\mb V_l$ defines the impurity potential matrix:
  
%16
\begin{equation}
\label{eq16}
{\mb V}_l  = \left(\begin{array}{cc}
\varepsilon_l & 0 \\
0 & -\varepsilon_l
\end{array}
\right)~.
\end{equation}

To the lowest order in $\varepsilon_i$ we get:
%17
\begin{equation}
\label{eq17}
\mb G(i,j;\omega) = \mb G^0(i,j;\omega)+\sum_l \mb G^0(i,l;\omega)\mb
V_l\mb
G^0(l,j;\omega) 
\end{equation}

Following Eqs. (\ref{eq5},\ref{eq17}) we  express interesting
quantities $\Delta_{ij}$ and $\Delta_{il}$\cite{Lit01b,Lit01d}, where  $l$
and
$j$ are
nearest 
neighbours of $i$, in the lowest
order
of  $\varepsilon_i$ perturbations by means of disordered Green function
(Eq. \ref{eq17}) and calculate the mean square deviation  of the paring
parameter as:
%18
\begin{equation}
\label{eq18}
 <|\delta \Delta_{ij}|^2 >=
<|\Delta_{ij}|^2>-|<\Delta_{ij}>|^2 ,
\end{equation}
where  $i$ and $j$ are nearest neighbour lattice sites.

The assumption is that random site energies $\varepsilon_i$ in Eqs.
(\ref{eq5},\ref{eq17},\ref{eq18}) are
independent variables then averages
$<\varepsilon_i>$ ,$<\varepsilon_i>$ being independent of the site
index $i$ and $<\varepsilon_i \varepsilon_j> = < \varepsilon_i^2 > \delta_{ij}$
lead to 

%19
\begin{equation}
< \delta \Delta_{ij} \delta \Delta_{ij}^*>
= \Gamma_{ij} < \varepsilon_i^2 > \nonumber \\
\label{eq19}  
\end{equation}

Finally, we calculate the coefficient
$\Gamma_{ij}$\cite{Gyo97,Lit01b}:

%20
\begin{equation}
\label{eq20}
\Gamma_{ij} =  \frac{1}{N} \sum_{\mb q}
\left|\frac{W_{ij}}{2N}
\sum_{\mb k}
\frac {\sub{\Delta}{k} \tilde{\epsilon}_{\mb k} + \sub{\Delta}{k} 
\tilde{\epsilon}_{\mb k - \mb q}} {(
E_{\mb k}+ E_{\mb k - \mb
q})
 E_{\mb k} E_{\mb k - \mb q}}{\rm e}^{\imath(\mb R_i -\mb R_j)
\mb k}
\right|^2~,
\end{equation}

Having solved the gap equations (Eqs. \ref{eq9}-\ref{eq11}) for the clean
system , we calculated
$\Gamma$ (Eq. \ref{eq20}) for each of
solutions we obtained. They are shown in each of Fig. 4, where we
plotted $\Gamma$ versus band filling $n$.
Analyzing the results, first
of all, we have to report that
for all cases $\Gamma$ has  small value (of order $\sim 10^{-7}$).
This implies that 
fluctuations of $\Delta_{ij}$ are  small in this system
and it is consistent with relatively large coherence length in this 
superconductor\cite{Gyo97}. 
Interestingly,
a real type of solution has the smallest value of $\Gamma$. This could 
mean
that 
this solution is favorable by disorder. 
Note that the maximum of $\Gamma$ (Fig. 4), however close to the van Hove
singularity,
occurs 
for a lower band filing $n$ than the maximum of the order parameter 
$\Delta_x$ (Fig. 2a).

\section{Conclusions and Discussion}

We have analyzed the effect of weak disorder on $p$--wave superconductor
in context of newly discovered superconductor  Sr$_2$RuO$_4$\cite{Mae01,Agt97}.
The  order parameter structure in this compound is still under 
debate\cite{Mae01,Ere01,Zhi01,Lit01c,Ann01}.

Fitting our system parameters to its $\gamma$ band structure we asked about
the stability of various solution in presence of disorder.
Note that the standard Abricosov-Gorkov formalism\cite{Abr59} 
applied for site
diagonal
nonmagnetic
impurities could give us only partial answer  on the disorder
effect on
superconductivity in Sr$_2$RuO$_4$.
It enables us to predict how strongly    
the critical temperature $T_c$ is
suppresed\cite{Mac98,Lit01a} but it is not
useful to  
analyze the stability of different solutions with the same $T_c$ appeared
in our case.   

Our preliminary results indicate that the complex solution has
the global minimum of free energy $F$ but it is the real one, with the
line
nodes, which  is favored
by 
disorder.
That result was obtained in one band model in
the lowest order
of perturbation
method and should be investigated further by a more sophisticated method 
which can treat a finite disorder,
and by considering  more realistic three dimensional, three orbital
electronic
structure of
Sr$_2$RuO$_4$\cite{Ann01}.

\section*{Acknowledgements}
This work has been partially supported by KBN grant No. 5P03B00221 and 
 the
Royal Society Joint Project.

\end{document}